\documentclass[sigconf,natbib=false]{acmart}
\AtBeginDocument{%
  }

\setcopyright{cc}
\setcctype[4.0]{by}

\acmYear{2025}\copyrightyear{2025}
\acmDOI{}\acmISBN{}
\acmConference[EARL@RecSys'25]{EARL Workshop at RecSys'25}{September 26, 2025}{Prague, Czech Republic}

\RequirePackage[
  datamodel=acmdatamodel,
  style=acmnumeric,
  ]{biblatex}

\begin{document}

\title{DenseRec: Revisiting Dense Content Embeddings for Sequential Transformer-based Recommendation}

\author{Jan Malte Lichtenberg}
\email{malte@usealbatross.ai}
\affiliation{%
  \institution{Albatross AI}
  \country{}
}

\author{Antonio De Candia}
\email{antonio@usealbatross.ai}
\affiliation{%
  \institution{Albatross AI}
  \country{}
}

\author{Matteo Ruffini}
\email{matteo@usealbatross.ai}
\affiliation{%
  \institution{Albatross AI}
  \country{}
}

\renewcommand{\shortauthors}{Lichtenberg et al.}

\begin{abstract}
Transformer-based sequential recommenders, such as SASRec or BERT4Rec, typically rely solely 
on learned item ID embeddings, making them vulnerable to the item cold-start problem, particularly 
in environments with dynamic item catalogs. While dense content embeddings from pre-trained models 
offer potential solutions, direct integration into transformer-based recommenders has consistently 
underperformed compared to ID-only approaches. We revisit this integration challenge and propose 
\textit{DenseRec}, a simple yet effective method that introduces a dual-path embedding approach. 
DenseRec learns a linear projection from the dense embedding space into the ID embedding space 
during training, enabling seamless generalization to previously unseen items without requiring 
specialized embedding models or complex infrastructure. In experiments on three real-world 
datasets, we find DenseRec to consistently outperform an ID-only SASRec baseline, even without 
additional hyperparameter tuning and while using compact embedding models. Our analysis suggests 
improvements primarily arise from better sequence representations in the presence of unseen 
items, positioning DenseRec as a practical and robust solution for cold-start sequential recommendation.
\end{abstract}


\keywords{Sequential Recommendation, Transformers, SASRec, Item Cold Start, Content Embeddings, Generalization}


\maketitle


\section{Introduction}
ID-based transformer architectures for sequential recommendation (SASRec~\cite{kang2018self}, BERT4Rec~\cite{sun2019bert4rec}, etc.) have achieved 
significant advances in recommendation and personalization quality in recent years. However, item cold-start 
remains a persistent challenge, particularly for real-world recommender systems operating with highly dynamic 
item catalogs---such as second-hand marketplaces with millions of new, unique items added daily, or short-video 
platforms with continuously emerging content. A natural approach to address this challenge is leveraging item 
content descriptions or images to enable generalization to previously unseen items. With pre-trained large language models and image embedding models~\cite{reimers-2019-sentence-bert, cherti2023reproducible} now widely available, practitioners can readily obtain dense content embeddings for newly added items. However, directly integrating these dense content embeddings into transformer-based sequential recommenders has proven challenging.

Previous attempts at using dense content embeddings directly as input to sequential transformers have consistently 
underperformed compared to pure ID-based approaches~\cite{singh2024better, zhang2024id, hou2024bridging, hou2023learning}. 
One fundamental issue is that while dense embeddings from pre-trained embedding models excel at capturing semantic 
content similarity, they fail to distinguish between semantically similar items with vastly different user appeal 
or contextual relevance~\cite{li2020sentence, zhang2024id} and thus struggle to learn item-specific popularity patterns (memorization)~\cite{singh2024better}. 

To address these limitations, recent work has explored alternative approaches including training specialized 
dense embedding models for recommendation tasks~\cite{zhang2024id, hou2024bridging} and quantization-based methods that create "semantic IDs" 
from content embeddings~\cite{rajput2023recommender, hou2023learning, singh2024better, yang2024unifying}. While promising, these approaches either require extensive pre-training of embedding models or add considerable infrastructural complexity through building and training a separate vector quantization model and generative retrieval mechanisms that necessitate changes to existing retrieval infrastructure. Indeed, several recent studies on industry-scale recommender systems continue to rely on ID-based sequential models~\cite{chen2025pinfm, khrylchenko2025scaling}.
From a practitioner's perspective, there is a clear need for approaches that can leverage readily available 
off-the-shelf embedding models without requiring specialized pre-training or complex architectural modifications 
to existing retrieval systems. 

In this work, we revisit the direct integration of dense content embeddings into sequential transformers and 
propose \textit{DenseRec}, a simple yet effective approach that addresses the fundamental limitations of naïve dense 
embedding integration. Our method learns a linear projection from the dense content embedding space into the ID 
embedding space \textit{during} training, enabling the model to leverage both semantic content information and 
collaborative signals. This dual-path training strategy allows generalization to previously unseen items while 
maintaining the ability to learn item-specific patterns from interaction data. Importantly, at inference time, our 
approach enables controllable selection between the two pathways: ID-based embeddings for known items (leveraging 
learned collaborative patterns) and content-based embeddings for cold-start items (enabling immediate generalization). 

In our experiments, we observe that this flexibility allows the model to exploit both representation types 
without compromising performance on either known or unseen items. Furthermore, our approach requires minimal architectural 
modifications to existing transformer-based recommenders and introduces only a single additional hyperparameter, which we find to be effective over a wide range of values and robust across datasets.

\section{Related Work}

\textbf{Sequential Recommendation}. Sequential recommendation has evolved significantly with the adoption of deep learning architectures. 
Early approaches used recurrent neural networks \cite{hidasi2015session, tan2016improved} to model user behavior sequences. 
The introduction of attention mechanisms led to more sophisticated models like BERT4Rec \cite{sun2019bert4rec} and SASRec \cite{kang2018self}, which 
apply self-attention to capture long-range dependencies in user sequences.

These transformer-based approaches have become the foundation for modern sequential recommendation systems due to 
their ability to capture complex sequential patterns and their superior performance on standard benchmarks. 
However, these early models rely entirely on learned ID embeddings from items present in the training set, making them vulnerable to the item cold-start problem or cross-domain recommendation settings when new items are introduced to the system.\newline

\noindent\textbf{Item Cold-Start with Content Information}. Traditional approaches to the item cold-start problem have 
relied on content-based filtering \cite{pazzani2007content} and hybrid methods \cite{burke2002hybrid} that combine 
collaborative and content signals, for example, by presenting items from both sources in different collections~\cite{gomez2015netflix} or blending them into a single ranking~\cite{lichtenberg2024ranking}. Early neural approaches explored auto-encoders for joint collaborative-content representations \cite{wang2015collaborative}, matrix factorization with item features \cite{gantner2010learning}, 
and multi-task learning for recommendation and content prediction \cite{bansal2016ask}. 
DropoutNet \cite{volkovs2017dropoutnet} addressed cold-start through content features and 
dropout-based training simulation, while neural collaborative filtering was extended with 
feature interaction layers \cite{he2017neural}. Recent work has also explored integrating dense embeddings from 
pre-trained language models in shallow auto encoders~\cite{vanvcura2024beeformer}.\newline

\noindent\textbf{Dense Embeddings in Transformer-based Recommenders}. Using dense embeddings from pre-trained 
embedding models as direct replacements for ID-based embeddings has been a common baseline in various studies, 
but these approaches generally underperform compared to pure ID-based methods \cite{singh2024better, zhang2024id, hou2024bridging}. 
Various works~\cite{hou2022towards, hou2024bridging, zhang2024id} have thus proposed enhanced mechanisms to pre-train 
new language embedding models on the recommendation data and task itself---as opposed to using off-the-shelf pre-trained 
models, which is the focus of our work. In a recent study on an industry-scale sequential recommender system, content embeddings were used to initialize the ID-based embedding table but then continued to be trained~\cite{celikik2024building}. This strategy only allows inference on cold-start items if the training does not change the geometry of the embedding space. Other works have propose to directly transform the text output of pre-trained generative large language models (LLMs) into recommendations~\cite{lichtenberg2024large, zhao2024recommender, lin2025can}, which has obvious limitations in the item cold-start scenario. \newline

\noindent\textbf{Semantic ID Approaches}. 
A recent line of work has proposed using "semantic IDs" 
to bridge the gap between content and collaborative filtering. These approaches typically 
employ vector quantization techniques to convert dense content embeddings into discrete 
token "codes" that can then be processed by transformer architectures.

TIGER \cite{rajput2023recommender} introduces a two-stage approach where 
RQ-VAE~\cite{lee2022autoregressive} (Residual Quantized Variational Autoencoder) is first trained to learn semantic item IDs from 
content, followed by a transformer that operates on these discrete representations. TIGER and similar 
approaches \cite{singh2024better, li2023text, hou2023learning, petrov2024recjpq} have shown promising results but
come with additional challenges including codebook collapse~\cite{zhu2024cost, zhai2025simple} and 
the need of training and maintaining the additional quantization model (including multiple additional hyper parameters 
including code length and codebook size).  

In these semantic ID approaches, each item is represented by a sequence or code of tokens, 
which fundamentally alters the recommendation retrieval process. This representation creates  
infrastructural challenges, as one cannot directly generate
a user representation from item interaction sequences and perform standard (approximate) nearest 
neighbor search in the dense embedding space~\cite{yang2024unifying, singh2024better}.\newline

\noindent\textbf{Our Contribution}. In contrast to existing approaches, DenseRec provides 
a simple yet effective solution that requires minimal architectural changes to existing 
transformer-based recommenders. Unlike naive dense integration approaches that typically underperform, 
our dual-path training strategy enables effective utilization of both collaborative and content signals. 
This positions our work as a practical middle ground that achieves strong cold-start performance 
without the complexity overhead of more sophisticated approaches.

\section{Method: DenseRec}
\begin{figure*}[t]
  \centering
  \includegraphics[width=\linewidth]{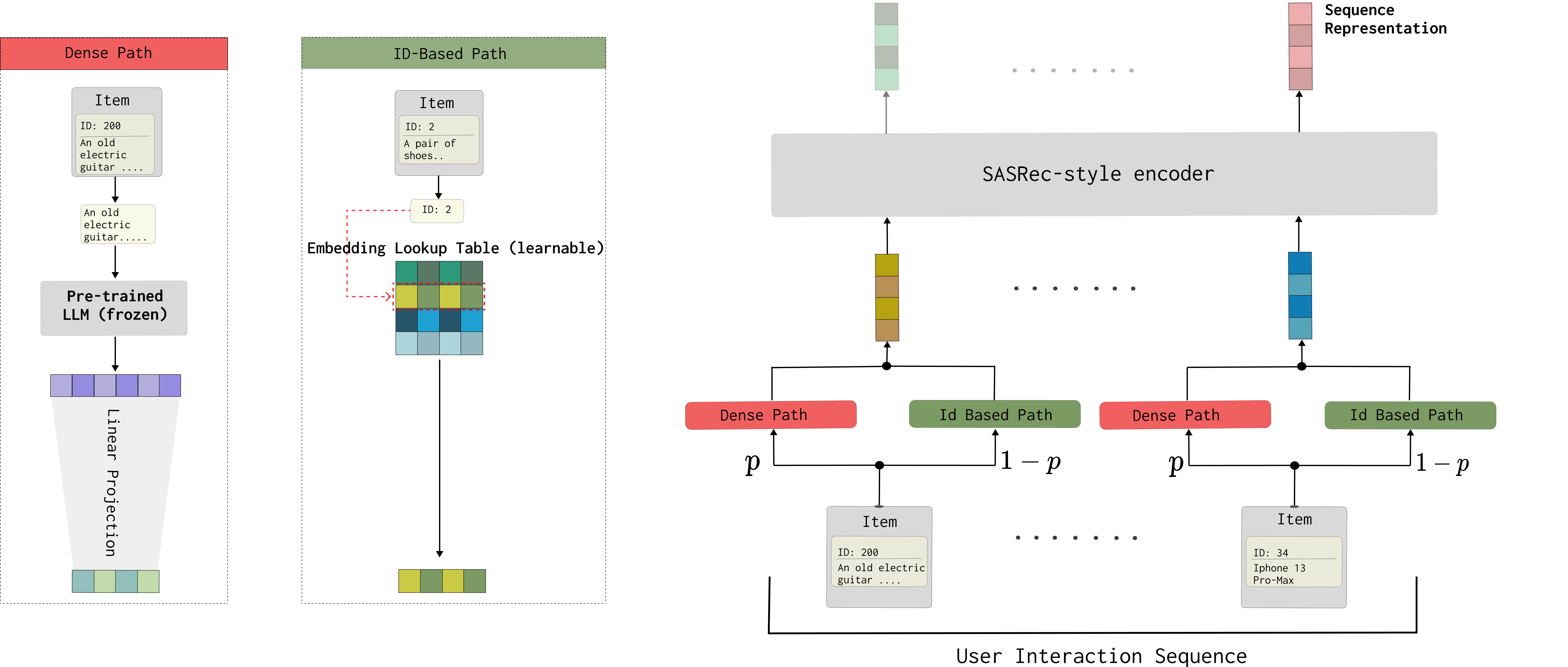}
  \caption{DenseRec architecture overview. The model maintains two parallel embedding pathways: 
  (1) ID Path using traditional learnable embeddings, and (2) Content Path using pre-computed content 
  embeddings projected into the ID embedding space via a learnable projection layer. During training, 
  a probabilistic selection mechanism determines which path to use for each token position.}
  \label{fig:DenseRec_architecture}
  \Description{Architecture diagram showing the dual-path design of DenseRec with ID embeddings 
  and dense content embeddings being processed through separate pathways before entering the 
  transformer layers.}
\end{figure*}

\subsection{Problem Formulation}

We consider the sequential recommendation problem where users interact with items over time. 
Let $\mathcal{U} = \{u_1, u_2, \ldots, u_{|\mathcal{U}|}\}$ be the set of users 
and $\mathcal{I} = \{i_1, i_2, \ldots, i_{|\mathcal{I}|}\}$ be the set of items observed during training. 
For each user $u$, we have a sequence of 
interactions $S^u = [i_1^u, i_2^u, \ldots, i_{|S^u|}^u]$ where $i_j^u \in \mathcal{I}$ represents 
the $j$-th item interacted with by user $u$. In a common production setting, the model is usually 
trained on sequences collected until a certain time stamp $t_0$ and the main task then is to p
redict a user's next item interaction given a sequence of items collected \textit{after} $t_0$. 

The key challenge we address is the \textit{item cold-start problem}: at test time, we may encounter 
items $i \notin \mathcal{I}$ that were not observed during training. For each 
item $i \in \mathcal{I} \cup \mathcal{I}_{new}$ (where $\mathcal{I}_{new}$ represents new items), 
we assume access to a dense content embedding $\mathbf{c}_i \in \mathbb{R}^{d_c}$ derived from 
pre-trained models applied to item descriptions, images, or other content features.

\subsection{DenseRec Architecture}

We formulate the DenseRec model as an extension to the SASRec transformer architecture with a dual-path design that can 
leverage both learned ID embeddings and dense content embeddings. The model maintains two parallel 
embedding pathways:
\begin{enumerate}
\item \textbf{ID Path}: Traditional learnable embeddings $\mathbf{E}^{id} \in \mathbb{R}^{|\mathcal{I}| \times d}$ where $d$ is the embedding dimension
\item \textbf{Dense Path}: Pre-computed content embeddings $\mathbf{C} \in \mathbb{R}^{|\mathcal{I}| \times d_c}$ projected into the ID embedding space, via a learnable projection layer $\mathbf{P}: \mathbb{R}^{d_c} \rightarrow \mathbb{R}^{d}$.
\end{enumerate}

Figure~\ref{fig:DenseRec_architecture} illustrates the overall architecture of our DenseRec model, 
showing how the dual-path design integrates both ID-based and content-based embeddings through the 
learned projection mechanism.
In our experiments shown below, we implemented $\mathbf{P}$ as a simple single linear transformation:
\begin{equation}
\mathbf{P}(\mathbf{c}_i) = \mathbf{W}_p \mathbf{c}_i + \mathbf{b}_p
\end{equation}
but future work could explore more complex, non-linear architectures.

\subsection{Dual-Path Training Strategy}

During training, DenseRec employs a probabilistic path selection mechanism controlled by the 
hyperparameter \textit{dense path probability}, denoted by $p_{dense} \in [0,1]$. For each token 
position in the input sequence, we randomly decide whether to use the ID path or dense path:
\begin{equation}
\mathbf{e}_i^{(t)} = \begin{cases}
\mathbf{E}^{id}[i] & \text{with probability } 1 - p_{dense} \\
\mathbf{P}(\mathbf{c}_i) & \text{with probability } p_{dense}
\end{cases}
\end{equation}
where $\mathbf{e}_i^{(t)}$ is the embedding used for item $i$ at position $t$ in the sequence.

This stochastic training strategy serves two purposes:
\begin{enumerate}
\item It forces the model to learn meaningful ID embeddings while simultaneously training the projection layer
\item It ensures the projection layer learns to map content embeddings to representations that are compatible with the transformer's learned dynamics
\end{enumerate}

The same probabilistic selection applies to the output embeddings used in the loss computation, 
ensuring consistency between input and output representations.

\subsection{Model Forward Pass}

Given an input sequence $S = [i_1, i_2, \ldots, i_n]$, the forward pass proceeds as follows:

\begin{enumerate}
\item For each position $t$, determine path selection $z_t \sim \text{Bernoulli}(p_{dense})$
\item Compute token embeddings:
\begin{equation}
\mathbf{h}_t^{(0)} = \begin{cases}
\mathbf{E}^{id}[i_t] & \text{if } z_t = 0 \\
\mathbf{P}(\mathbf{c}_{i_t}) & \text{if } z_t = 1
\end{cases}
\end{equation}
\item Apply positional embeddings and transformer layers as in standard SASRec:
\begin{equation}
\mathbf{H}^{(l+1)} = \text{TransformerBlock}^{(l)}(\mathbf{H}^{(l)})
\end{equation}
\item Generate final sequence representation $\mathbf{H}^{(L)} \in \mathbb{R}^{d}$ from the output of the last item in the sequence.
\end{enumerate}

\subsection{Loss Function and Training}

We use the same loss function as SASRec with negative sampling, but apply the dual-path strategy 
to both input sequences and target items. For a sequence ending with target item $i_{target}$ and 
negative samples $\{i_{neg}^{(j)}\}_{j=1}^K$, we compute:

\begin{equation}
\mathcal{L} = -\log \sigma(\mathbf{h}_n^T \mathbf{e}_{i_{target}}) - \sum_{j=1}^K \log \sigma(-\mathbf{h}_n^T \mathbf{e}_{i_{neg}^{(j)}})
\end{equation}

where $\mathbf{h}_n$ is the sequence representation and $\mathbf{e}_{i}$ is the output embedding for 
item $i$, which is either the ID-based embedding or the projected dense embedding, selected using 
the same probabilistic mechanism as with the input embeddings. The model 
parameters $\{\mathbf{E}^{id}, \mathbf{W}_p, \text{Transformer weights}\}$ are jointly optimized 
using standard backpropagation.

\subsection{Inference and Handling of Cold-Start Items}

At inference time, DenseRec follows the standard sequential recommendation inference process 
similar to SASRec and other transformer-based approaches. Given a test sequence, the model 
generates a sequence representation $\mathbf{h}_n$ from the final position. This representation 
is then used to compute similarity scores with all candidate items via dot-product operations. 
Retrieval can be performed using k-nearest neighbor (KNN) or approximate nearest neighbor 
(ANN) methods for large item catalogs to generate the final recommendations. DenseRec handles 
different item types during candidate scoring as follows:

\textbf{Known Items}: For items $i \in \mathcal{I}$ that were observed during training, we 
exclusively use the learned ID embeddings $\mathbf{e}_i = \mathbf{E}^{id}[i]$.

\textbf{Cold-Start Items}: For items $i \notin \mathcal{I}$ that were not seen during training, 
we exclusively use the dense path $\mathbf{e}_i = \mathbf{P}(\mathbf{c}_i)$. This allows the model 
to generate meaningful representations for new items without requiring retraining, leveraging 
the projection layer learned during the dual-path training process.

\textbf{Arbitrary Item Addition}: A practical advantage of our approach is that we can dynamically 
add arbitrary items to the candidate set as long as we have their dense content embeddings. 
New items can be immediately incorporated into the recommendation process by simply computing 
their projections $\mathbf{P}(\mathbf{c}_i)$ into the ID embedding space that is used by the KNN or ANN
method for candidate retrieval, enabling real-time 
catalog expansion without model retraining. This contrasts with hybrid approaches that require 
a catch-all ID for all cold-start items or semantic ID and generative retrieval approaches 
where performing retrieval on never-seen-before item code combinations can be challenging.

\section{Experimental Setup}

Our experimental evaluation focuses on assessing whether adding DenseRec's dual-path mechanism 
to an existing architecture (SASRec) can improve performance in a cold-start setting with minimal 
additional effort---specifically 
without additional hyperparameter optimization and little engineering overhead.

\subsection{Datasets}

We evaluate our approach on three categories from the Amazon Reviews 2023 
dataset \cite{hou2024bridging}: \textit{Sports and Outdoors} (\textbf{Sports}), \textit{Toys and Games} (\textbf{Toys}), and \textit{Video Games} (\textbf{Video}). 
These categories were selected for consistency with related work on semantic ID approaches \cite{rajput2023recommender} and 
content-based sequential recommendation \cite{hou2024bridging}.

To ensure our experimental setup closely mirrors real-world production scenarios, we adopted 
the \textit{absolute-timestamp splitting} methodology proposed by the Amazon Reviews 2023 data set 
authors~\cite{hou2024bridging}.\footnote{See also \url{https://amazon-reviews-2023.github.io/data_processing/0core.html\#absolute-timestamp-splitting}} This approach 
splits data based on interaction timestamps rather than the commonly used leave-one-out methodology, 
creating temporally coherent train/validation/test sets. In particular, unlike leave-one-out splitting where large portions of test sequences are observed 
during training, absolute timestamp splitting ensures that test sequences can contain entirely 
new users or interactions that occurred after the training cutoff, more accurately simulating 
the real-world production setting. For training we filter out all items with less than 5 interactions and 
sequences with fewer than 2 item reviews. We retained all test set items that were absent  
from the training set to simulate realistic item cold-start scenarios. 
This design choice reflects production environments such as second-hand marketplaces where new, 
previously unseen items are continuously added by sellers and must be recommended without 
historical interaction data.

Table~\ref{tab:dataset_stats} provides detailed statistics for each dataset, highlighting the cold-start challenges inherent in our experimental setup.

\begin{table}[h]
\centering
\caption{Dataset statistics for Amazon Reviews 2023 categories including the ratio of cold-start "target" items and ratio of cold-start items among all items that are used to generate user/sequence representations at test time.}
\label{tab:dataset_stats}
\begin{tabular}{lccc}
\toprule
\textbf{Statistic} & \textbf{Toys} &  \textbf{Sports}  & \textbf{Video} \\
\midrule
\# Items & 266,346 & 364,657 & 113,297 \\
\# Users & 2,168,966 & 2,787,852 & 620,055 \\
Avg. sequence length & 5.72 & 5.62 & 6.18 \\
\midrule
Cold-start target items & 49.2\% & 36.8\% & 51.7\% \\
Cold-start items in test seqs. & 24.7\% & 21.9\% & 23.0\% \\
\bottomrule
\end{tabular}
\end{table}

\subsection{Models}

The goal of the main experiment was to demonstrate the added value of the dense-to-ID projection in cold-start scenarios. 
We therefore compared \textbf{DenseRec} directly to the standard \textbf{ID-based SASRec} as described in 
Kang et al.~\cite{kang2018self}. While various improvements to the original SASRec architecture have been 
proposed (e.g., improved negative sampling strategies~\cite{petrov2023gsasrec}, or modifications to the 
loss function~\cite{wilm2023scaling, abbattista2024enhancing}), many of those extensions are 
orthogonal to---and could potentially be combined with---the DenseRec architecture. 

To demonstrate the practical simplicity of our approach, we employed the following hyperparameter optimization (HPO) strategy that provides an unfair advantage to the baseline ID-based SASRec model: We performed HPO for the ID-based SASRec baseline to establish a suitable model configuration across all data sets. These optimized hyperparameters were then directly transferred to DenseRec with a single addition: a fixed $p_{dense}$ of $0.5$ across all datasets (that is, a fair "coin flip" split between dense and ID path for every item, see also the discussion in Section \ref{sec:results} on why this middle-ground makes sense intuitively). We thus selected the best parameters for the baseline model and performed \textit{no} dataset-specific or model-specific HPO for DenseRec itself. This allows us to have confidence that the obtained results were not due to a more intensive HPO on our method than on the baseline method, highlighting the robustness and ease of deployment of the proposed approach. The complete hyperparameter specifications and exact HPO method for all models are provided in the Appendix.

\subsection{Content Embeddings}

For dense content embeddings, we used the \texttt{all-MiniLM-L6-v2} model\footnote{\url{https://huggingface.co/sentence-transformers/all-MiniLM-L6-v2}} from the \texttt{sentence-transformers} library ~\cite{reimers-2019-sentence-bert} to embed item content. Text inputs were formatted as 
\begin{center}
\texttt{"title: \{title\}, description: \{description\}"}
\end{center}
and capped at 300 characters to ensure consistent processing across all categories. This, by today's standards, relatively lightweight text embedding model (22.7M parameters, embedding dimension of 384) was chosen to demonstrate that our approach works effectively even with compact content representations, making it practical for production deployment where computational efficiency is important.

\subsection{Evaluation Protocol}

All models were trained on the provided training splits and evaluated on the corresponding test sets. For each test sequence, we evaluated model performance by predicting only the last item in the sequence, using the preceding items as input context. Crucially, due to the absolute timestamp-based splitting methodology, no portion of any test sequence was observed during training, ensuring a truly out-of-time evaluation that reflects real-world deployment scenarios.

We used Hit Rate@100 as our primary evaluation metric, computed against the full item catalog (all items present in both training and test sets) rather than a random sample of negatives. This approach reduces evaluation variance and better reflects production recommendation scenarios where models must retrieve from the entire available inventory. The choice of k=100 aligns with modern retrieve-then-rank recommendation architectures where high recall in the retrieval stage is critical for subsequent re-ranking performance. 

For cold-start evaluation, we specifically analyze performance on test set items that were not observed during training, providing direct measurement of generalization capability to unseen items.

\begin{figure*}[t]
  \centering
  \includegraphics[width=0.32\linewidth]{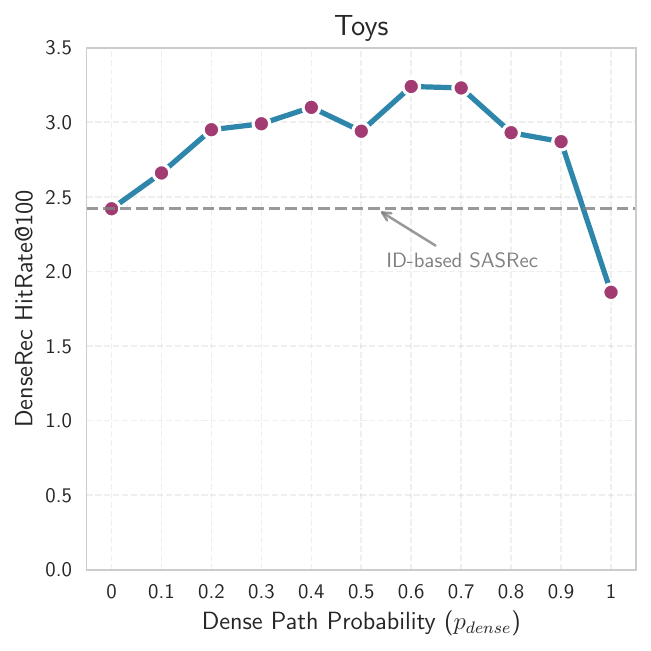}
    \includegraphics[width=0.32\linewidth]{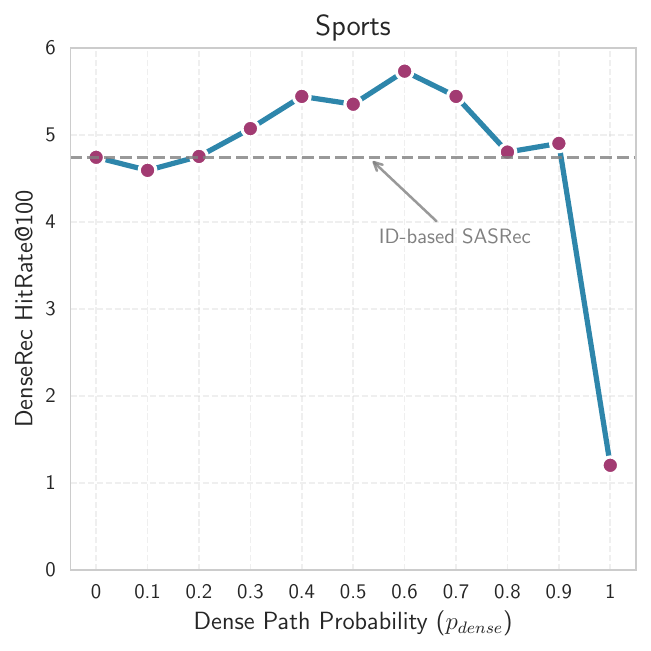}
      \includegraphics[width=0.32\linewidth]{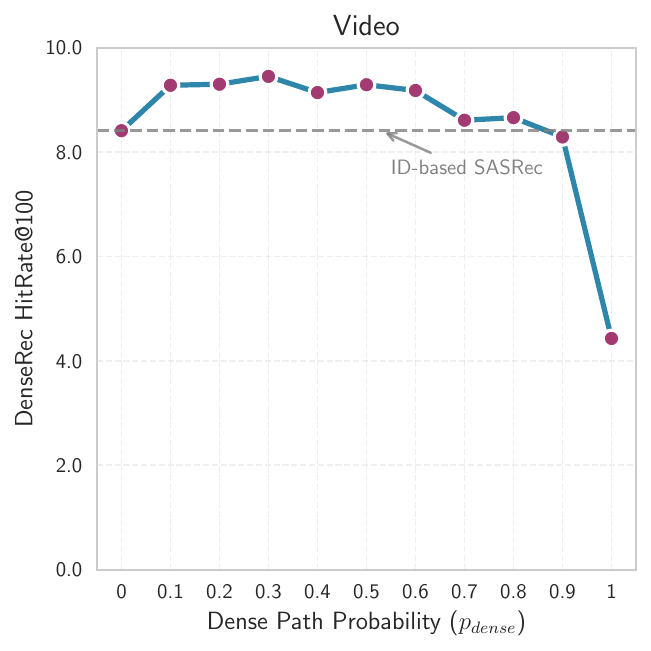}
  \caption{DenseRec performance as a function of the $p_{dense}$ parameter and the ID-based SASRec baseline (dashed line and $p_{dense} = 0.0$). For $p_{dense} = 1.0$, the model is equivalent to the "naïve" implementation of using only dense content embeddings.}
  \label{fig:dense_path_prob}
\end{figure*}

\section{Results} \label{sec:results}

\textbf{Overall Performance.} Table~\ref{tab:main_results} presents the Hit Rate@100 test set performance comparison between our DenseRec approach and the ID-based SASRec baseline. 
\begin{table}[h]
\centering
\caption{Hit Rate@100 performance comparison and relative improvement of DenseRec compared to ID-based SASRec across all three data sets.}
\label{tab:main_results}
\begin{tabular}{lccc}
\toprule
\textbf{Model} & \textbf{Toys} & \textbf{Sports} & \textbf{Video} \\
\midrule
ID-based SASRec & 2.42 & 4.75 & 8.41 \\
DenseRec (Ours) & \textbf{3.25} & \textbf{5.35} & \textbf{9.37} \\
\midrule
Relative Improvement & +34.3\% & +12.6\% & +11.4\% \\
\midrule
\% of cold-start items among hits & 2.4\% & 0.4\% & 2.3\% \\
\bottomrule
\end{tabular}
\end{table}

DenseRec consistently outperformed the ID-based approach across all three categories, 
demonstrating the effectiveness of our dual-path training strategy and learned projection mechanism.
We note again we used a small backbone embedding model and no DenseRec-specific HPO, 
suggesting that there is headroom for further improvements compared to the ID-only SASRec.\newline

\noindent \textbf{Where does the performance lift come from?}
The DenseRec model can leverage cold-start embeddings in two ways to provide a lift compared to the ID-only model:
\begin{enumerate}
\item \textit{Cold-start items as target}: By projecting cold-start items into the retrieval candidate space, 
DenseRec can retrieve relevant items that were never observed during training.
\item \textit{Cold-start items in the sequence}: Even if the target item is a known item, DenseRec might 
be able to build better (test) sequence representations by including cold-start items, whereas the ID-based model
has to exclude those items when building the user representation.
\end{enumerate} 
Table~\ref{tab:dataset_stats} shows the prevalence of cold-start items in the test sequences and the test target items.
Table~\ref{tab:main_results} shows the percentage of hits (that is, correct predictions) that were on cold-start items. These range from only 0.4\% (Sports) to 2.4\% (Toys), suggesting that DenseRec obtained its superior overall prediction performance by leveraging cold-start items to build better sequence representations as opposed to correctly predicting cold-start items.\newline

\noindent\textbf{The impact of $p_{dense}$}. The dense path probability parameter $p_{dense}$ determines 
how often the dense embedding path is used instead of the original ID-based path. 
Intuitively, the parameter thus determines how much of the training data is used to 
learn the ID-embedding space vs. learning the projection from the dense into the ID-embedding space. 
It also seems intuitively reasonable to avoid either extreme: for $p_{dense} = 0$, the model 
will not use train the projection layer at all and thus cannot utilize the dense content embeddings, 
which is basically equivalent to using ID-based recommender systems. For $p_{dense} = 1$, the ID-based embedding 
space will be learned solely through via projected dense embeddings, similar to the direct dense 
embedding implementations observed in existing works. In our main experiment described above, 
we therefore opted for $p_{dense} = 0.5$ as a happy medium that intuitively makes sense and 
does not require additional HPO. 

In the following set of experiments, we computed the performance of the DenseRec model 
for $p_{dense} \in [0.0, 0.1, \dots, 0.9, 1.0]$ across all three data sets to evaluate the robustness of 
the model's performance with respect to the choice of $p_{dense}$. Figure \ref{fig:dense_path_prob} shows 
HitRate@100 for the different values of $p_{dense}$ as well as the performance of the ID-based SASRec model
from Table~\ref{tab:main_results}, for all three data sets. Overall, the DenseRec model is remarkably robust 
to the choice of $p_{dense}$ anywhere except for the extreme value of $p_{dense}=1.0$, matching or outperforming the
ID-based model across all values between $0.2$ and $0.8$. Indeed, for Toys and Sports, we observe a (quite noisy)
inverse U-shape in the performance, which roughly matches the intuition to avoid either extreme of $0.0$ or $1.0$ 
described above. That said, our "intuitive" choice of $p_{dense} = 0.5$ turned out
to never be the best performing value, indicating that an additional HPO on the $p_{dense}$ might further improve the performance of 
DenseRec. 

For Video, we do not observe a U-shape, but an almost monotonic decrease in performance with increasing 
$p_{dense} \geq 0.1$, which might indicate that text embeddings were less useful for this data set.

\section{Conclusion}

Combining behavioral signals with content information is among the earliest and most enduring concepts in recommender systems research~\cite{burke2002hybrid}. 
In this paper, we revisited this fundamental idea within the specific context of transformer-based sequential recommender models. 
We began by highlighting that directly integrating dense content embeddings into transformer-based sequential recommenders 
has consistently underperformed compared to purely ID-based methods~\cite{singh2024better, zhang2024id, hou2024bridging, hou2023learning}, and we replicated these findings in our experiments, where we observed low hit rate values for $p_{dense} = 1.0$ (which is equivalent to using only content embeddings).

To address this limitation, we introduced \textit{DenseRec}, a simple yet effective approach designed explicitly for ease of integration. 
DenseRec involves minimal architectural modifications to existing ID-based sequential recommenders and introduces only a 
single additional hyperparameter, thus significantly reducing complexity, hyperparameter optimization requirements, and infrastructural overhead.

Our experimental design was intentionally favoring the baseline method: Hyperparameter optimization was performed exclusively for the 
ID-based SASRec model, whereas DenseRec was evaluated without any dedicated tuning and used only a modestly sized pre-trained 
language model for content embeddings. Despite these constraints, DenseRec consistently outperformed the ID-only baseline, 
demonstrating its practical effectiveness and robustness.

The classical item cold-start problem focuses primarily on predicting items unseen during training. However, modern recommender 
systems increasingly require real-time responsiveness, continuously integrating new items and user interactions into updated 
recommendations. This real-time necessity makes the cold-start challenge worse because purely ID-based methods 
inherently lack mechanisms to incorporate newly introduced items into user or sequence representations dynamically.

Our findings suggest that DenseRec represents a promising step toward resolving this  variant of the cold-start problem. 
The observed performance gains were largely attributable to DenseRec's improved capability to construct meaningful sequence 
representations from test sequences containing previously unseen items.

In the current work, we used the SASRec architecture as our backbone model for sequential 
transformer-based recommendation, given its simplicity, flexibility, and consistently 
strong, near state-of-the-art performance~\cite{petrov2022systematic, petrov2023gsasrec}. However, the DenseRec approach is not limited 
to SASRec and can readily be integrated into extended models such as gSASRec~\cite{petrov2023gsasrec} 
or other ID-based sequential recommenders, including more recent architectures like 
Mamba4Rec~\cite{liu2024mamba4rec}. The minimal architectural changes required by DenseRec 
make it broadly applicable across a wide range of existing transformer-based recommendation systems.

\begin{acks}
To Davide Abbattista, for providing feedback on a draft of this paper as well as to Johan Boissard, Kevin Kahn, and the entire team at Albatross AI, for making this work possible.
\end{acks}

\printbibliography

\appendix

\section{Appendix}

\subsection{Hyper parameter settings}

Table \ref{tab:hps} shows the parameters that we used for the baseline ID-SASRec implementation.

\begin{table}[h]
\centering
\caption{SASRec hyperparamters}
\label{tab:hps}
\begin{tabular}{ll}
\toprule
\textbf{Hyperparameter} & \textbf{Value} \\
\midrule
Embedding dimension          & 64 \\
Epochs                       & 20 \\
Batch size                   & 512 \\
Max sequence length          & 30 \\
\# attention heads    & 2 \\
\# transformer blocks & 3 \\
Dropout rate                 & 0.5 \\
Use positional embeddings    & True \\
Negative samples per positive & 64 \\
\bottomrule
\end{tabular}
\end{table}

As explained in the main text, for the DenseRec model we used the exact same hyperparameters and added a constant dense path probability of $p_{dense} = 0.5$.

\end{document}